# Decentralized Digital Currency System using Merkle Hash Trees

Shreekanth M Prabhu, Natarajan Subramanyam, Ms. Shreya P Krishnan and Ms. Brindavana Sachidananda,

## Abstract:

In India, post demonetization exercise in 2016, digital payments have become extremely popular. Among them, the volume of transactions using Paytm wallets and UPI (Unified Payment Interface) have grown manyfold. The lockdowns due to COVID-19 Pandemic have furthered this trend. Side by side, crypto-currencies such as bitcoin are also gaining traction. Many countries are considering issuing a Digital Currency via their Central Banks. In this paper, we propose a novel Decentralized Digital Currency System (DDCS) that makes use of Merkle Hash-Trees as Authenticated Data Structures. DDCS uses a Ledger-less, distributed, peer-to-peer architecture. We name the proposed currency δ-Money. δ-Money is intended as a replacement for physical currency and has in-built security features that rival crypto-currencies. Transactions using δ-Money happen in a disintermediated manner but with post-facto reconciliation. In place of Central Bank-issued Digital Currency (CBDC), we envisage a scenario where multiple Payment Banks issue digital currencies that have stable valuations without being subject to either volatility or perennial devaluation.

**Biographical Notes**

**Dr Shreekanth M Prabhu** is currently working as Professor and Head of the Department of Computer Science and Engineering at CMR institute of Technology, Bengaluru, India. His research interests include Social Networks, E-Governance and Linguistics.

**Dr Natarajan Subramanyam** is currently Professor in Computer Science and Engineering, Department of CSE of PES University, Bangalore, India. He has 140+ research articles in reputed Conferences and Journals. He has guided 7 Research Scholars towards PhD.

**Ms. Shreya P Krishnan**, from Bangalore, India, is currently working at Hewlett Packard Enterprise as n R&D software engineer. Her main areas of interest include databases, data analytics, machine learning, operating systems and blockchain.

**Ms. Brindavana Sachidanand** is currently working with Qualcomm Inc, USA as IT Software Engineer. Her areas of interests include Machine Learning, Data Analytics and Databases.

# 1. Introduction

The concept of money goes back millennia. Without money, people need to barter goods and services that require the double coincidence of wants. Money is thus a means of exchange and a measure of value whereas currency serves as the medium of exchange. However, in common parlance money and currency get used interchangeably. The main form of money we use now is fiat currency/money which has no intrinsic value but is guaranteed by the Government. Fiat money acts as a claim on the central bank. In the case of commodity money, which was historically widely used, the value is derived (in part or full) by the commodity used to create money such as gold and silver. The bank deposits/drafts and the like stand for a claim on commercial banks. Fiduciary media or money substitutes such as letters of credit are yet another mechanism to facilitate transactions. Two other functions money does are unit of accounts and store of value. To qualify as a unit of accounts money should be divisible into smaller units without losing value. To act as a store of value, money must be able to be reliably saved, stored, and retrieved. In this case, the value of the money must also remain stable over time. Further money can also function as a "standard of deferred payment," which means that its status as legal tender allows it to function for the discharge of debts".

Fiat currencies were traditionally linked with silver standards, gold standard or gold exchange standards, and so on. Srikant Krishnamachary [1] covers the entire history of the Indian monetary standard from the 6$^{th}$ Century BCE to modern times. He recounts the prevalence of gold, silver, and copper coins as a medium of exchange in India for thousands of years till India moved away from Gold Standard in the 1890s. Dr, Ambedkar [2] has recounted how Indian Rupee went through a tumultuous period during British rule and how India lost out in the bargain relative to the colonial power. In summary over the past couple of centuries money which was based on commodities gave way to Fiat Money in the form of physical currency with no intrinsic value.

In the last couple of decades, the usage of fiat currencies in physical form has come down due to the increasing prevalence of digital payments. Digital mode has advantages as physical currencies carry the risk of counterfeiting and loss due to theft. For the Government tracking the use of physical currencies for illicit activities is a big challenge. The cost of printing the currencies is also substantial.

In India, cash and cheques increasingly have given way to the use of debit/credit cards followed by inter-bank transfers using NEFT/IMPS and RTGS services. The modes of digital payment got further expanded with the use of wallets enabled by payment service providers and the launch of Unified Payment Service (UPI) by the National Payment Corporation of India (NPCI). With UPI the transfers directly happen to bank accounts by linking them with a virtual address. UPI services are widely used now for transferring funds between parties. However, there are many accounts of fraud reported periodically. When the frauds happen, the entire monies held in a bank account may be subject to the risk of potential debit. While inter-bank transfers were used generally for high-value transactions, UPI is used for low-value and high-volume transactions, where settlement happens at the individual transaction level. This leads to transaction failures due to network load.

Post-independence till 1991's, Indian Rupee was pegged at an artificially high value. After which Rupee has gone through perennial devaluation. Part of the reason here is the economic policies followed by the Government. Most Economists support Keynesian Economics where Governments intervene to manage the values of their currencies typically on the lower side. This can happen due to bail-outs of businesses that fail, over-spending on welfare, and spending to stimulate the economy. This point was broadly addressed by John Nash [3-4] in his proposal on ideal money where he particularly deplored the decadence of value of fiat money due to Keynesian Economic ideas. "Good money," Nash argued, is money that is expected to maintain its value over time. "Bad money" is expected to lose value over time. According to Grisham's law, bad money drives out the good. When it comes to Indian Rupee, Team PGurus [5] argues that Government is enriching exporters at the cost of taxpayers, by keeping the value of the Rupee artificially low. Frequent incidence of bad loans makes the matters worse. It was John Nash who was vocal about money outside the system of Governments, which currently goes by the name virtual currency. The blogger [6] links John Nash's ideal money with bitcoin. David Lee Kuo Chuen [7] has covered in an edited volume different facets of digital currency.

The Cryptocurrencies such as bitcoin are also finding traction in India. However, they are too volatile and bear too high a valuation (at least in the case of bitcoin). The policy stance of the Government is ambivalent towards Cryptocurrencies, even going to the extent of proposing a ban on them [8]. To counter the cryptocurrencies, the Government of India is also monitoring the situation as far issuing a digital Rupee which like fiat currencies has a claim on the central bank but available in digital form. U.S. nonprofit Digital Dollar Project is in the process of launching five pilot programs over the next 12 months to test the potential uses of a U.S. central bank digital currency, the first effort of its kind in the United States [9]. China has already launched Digital Yuan [10]. Many experts [11] see introducing a Digital Rupee as the next logical step for India considering its phenomenal achievement in the widespread adoption of digital payments.

In India, the Central Bank gave licenses to eleven Payment Banks in 2015. These banks were to practice narrow banking where deposits were capped to Rs 100,00 and no lending was allowed. Payment Banks were supposed to provide services like Pre-Paid Instruments (PPIs) – accounts where customers deposited money and used it for specific small payments – utilities, shopping, business bills – and so on. However, within few months Unified Payments Interface was launched in India in 2016, which got a lot of traction and affected the prospects for Payment Banks [12]. Payment Banks being supervised commercial banks remain an opportunity to be tapped in the digital currency ecosystem.

In this paper, we propose a novel Decentralized Digital Currency System (DDCS) where transactions are done using δ-Money (the name given to the proposed currency). δ-Money uses Merkle-Hash Trees to establish provenance and enable detection of any kind of tampering and B+

Trees to store balance change records. The transactions happen directly between peers in a disintermediated manner. The integrity of transactions as well as conservation of currency is validated, typically after every transaction. We envisage δ-Money being issued by Payment Banks. The DDCS and the methods used herein are disclosed in Indian Patent Application [13].

The rest of the paper is structured as follows. *Section 2, Literature Survey* concisely covers the literature associated with the emergence of digital currency/cryptocurrency. *Section 3, Decentralized Digital Currency System (DDCS)*, describes the proposed methodology. *Section 4, Discussions* covers the comparative analysis of the DDCS with contemporary systems and lists relative merits/limitations. *Section 5, India Considerations* makes a case for deploying DDCS in India. *Section 6, Conclusions* concludes the paper.

## 2. Literature Survey

To be most useful as money, a currency should be: 1) fungible, 2) durable, 3) portable, 4) recognizable, and 5) stable. The currency is fungible if units of the good(denominations) are of relatively uniform quality so that they are interchangeable with one another. It is durable if it retains its usefulness in future exchanges and be reused multiple times. The currency is portable if can be conveniently carried or transported. An indivisible good, immovable good, or good of low original use-value can create issues. Currency needs to be recognizable and differentiable from counterfeits or fakes. Further currency needs to have a stable valuation. Instability in valuation can cause friction in transactions. A lucid account of the functions of money is given here [14-16].

In 1983, a research paper by David Chaum introduced the idea of digital cash [17,18]. His motivation was to make the payments untraceable. He made use of "blind signatures" that prevent the bank from linking users to coins, providing non-traceability akin to cash. This was followed by many developments that facilitated payments using electronic money/internet money; payments to vending machines using mobile payments and peer-to-peer transactions launched by services such as PayPal in 1998. Here the digital was just the form and mode, money or currency transferred continued to be traditional and not virtual.

All these developments over the years led to the concept of digital money/currency being rather broad [19]. Digital currency is used interchangeably with digital money, electronic money, or electronic currency. It is available in digital form in contrast to physical such as banknotes and coins. It can be central bank-issued money accounted for in a computer database or digital base money which is a claim on the central bank the way notes in circulation have but issued in digital form. According to ECB [20], two important factors that may make digital base money attractive are (i) ability to have a claim on the central bank instead of commercial banks, as the latter may default, and (ii) availability of technologies such as Distributed Ledgers, which sooner or later will gain maturity

Alternatively, digital money can be virtual money that works across borders, without any backing from any of the Governments or just restricted to a close-knit community or an online gaming outlet. A virtual currency is so-called to indicate that it has no legal tender. A currency (typically virtual) that makes use of cryptographic algorithms is called cryptocurrency, the most famous among them is the bitcoin.

Digital currency is a money balance recorded electronically on a stored-value card or other devices or maintained in a network. Digital money can either be centralized, where there is a central point of control over the money supply, or decentralized, where the control over the money supply can come from various sources Another way to look at digital/electronic/network money is to compare it with deposits which have a claim on the bank. However, some central banks consider bank money recorded in computers as simply traditional money and only the mode of the transaction as digital [21].

In summary digital currency facilitates the exchange of value digitally. Digital Currency may be implemented using a debit card, credit card, or a wallet managed by a mobile phone application or a computer that moves money electronically. Another motivation to create currencies outside the Government system was the loss of value of a currency due to Government policies and reckless spending of public money. The third motivation for digital currency was anonymity, which was particularly addressed by Cryptocurrencies.

We look at the origin of Crypto-currency here. There was a set of free thinkers who believed that using cryptography we can make Governments pretty much unnecessary. This notion called Cryptographic Anarchy [22] inspired many technologists. Wei Dai [23] conceived digital money called B-money in an essay, where he talks about money which is transacted in anonymity where everyone uses pseudonyms. Wei Dai laid down a 5-step protocol that included (i) Creation of Money, (ii) Transfer of Money, (iii) Effecting of Contracts, (iv) Conclusion of Contracts, and (v) Enforcement of Contracts. Analysis of B-Money is given here [24]. The b-money proposal was acknowledged in the bitcoin proposal [25], the first major cryptocurrency that makes use of blockchain and became a household name. The virtual/cryptocurrencies such as bitcoin promise to do away with the role of third parties as well as Government

Bitcoin is a cryptocurrency, a form of electronic cash. Bitcoin can be sent from user to user on a peer-to-peer bitcoin network without the need for any intermediaries. Bitcoin creates money using a cryptographic technique called mining. Bitcoin makes use of Blockchain which securely stores transactions in public ledgers. Each transaction block is first represented as a Merkle Tree and only the resulting hash is stored in the ledger, which contains the chain of such hashes. The system is distributed in the sense that there is no particular coordinator. The transactions are confirmed/validated by miners, who download the blockchain and then approve the addition of new blocks. The value of bitcoins is subject to volatility.

Bitcoin was launched in 2009, which marked the start of decentralized blockchain-based cryptocurrencies with no central server, and no tangible assets held in reserve. As there was no central organization running the operations of cryptocurrencies, it was a challenge for Governments to regulate them.

Joseph Bonneau et al. [26] have covered the research perspectives and challenges for bitcoins and cryptocurrencies in a comprehensive manner. They consider "Bitcoin's three main technical components as transactions (including scripts), the consensus protocol, and the communication network. Bitcoin publishes transactions as matching outputs and inputs. Using cryptographic signatures, it is assured that only a valid recipient of a previous transaction referring it in a follow-up transaction. Bitcoin does not have the notion of users, accounts, and account balances. The authors [26] say "Arguably, there is little that is deeply innovative about Bitcoin's transaction format. However, the use of a scripting language to specify redemption criteria and the realization that transactions can specify the entire state of the system are non-obvious design choices given prior cryptocurrency systems, both of which have been standard in essentially all subsequent designs". Further to prevent double-spend attacks the state of blockchain transactions has to be agreed upon by all peers. "Bitcoin establishes consensus on the blockchain through a decentralized, pseudonymous protocol dubbed Nakamoto consensus. This can be considered Bitcoin's core innovation and perhaps the most crucial ingredient to its success. Any party can attempt to add to the chain by collecting a set of valid pending transactions and forming them into a block. The core ingredient is the use of a challenging computational puzzle (usually given the slight misnomer proof of work) to determine which party's block will be considered the next block in the chain." At any given time, the consensus blockchain is the "longest" version. If they are two equal-length chains, miners can choose either fork. Due to the random nature of the computational puzzle, one blockchain will eventually be extended further than the other at which point all miners should adopt it. Generally, transactions in a specific block are considered confirmed after 6 more blocks get added to the chain it is on.

Further, in bitcoin, there is no authority to induct peers into the network. Any node can join the network by connecting to a random sample of other nodes. Peers who join the network initially need a way to find out about other peers. This happens through diffusion over a well-connected with a low degree and low diameter.

Swan [27] presents 7 technology challenges for the adoption of Blockchain Technology in the future. These include Throughput, Latency, Size and Bandwidth, Security, Wasted Resources, Usability, Versioning, Hard Forks, and Multiple Chains.

Yli-Huumo et al. [28] have published a study in 2016 on the research topics, challenges, and future directions in blockchain Technology. The results of their study show that "focus in over 80% of the papers is on Bitcoin system and less than 20% deals with other Blockchain applications including e.g., smart contracts and licensing. The majority of research is focusing on revealing and improving limitations of Blockchain from privacy and security perspectives, but many of the

Page **6** of **37**

proposed solutions lack concrete evaluation on their effectiveness. Many other Blockchain scalability-related challenges including throughput and latency have been left unstudied".

An evaluation of bitcoin to serve as main-stream currency is done by Wharton Knowledge Series Paper [29] and the findings are not encouraging. According to the Wharton fellows "Bitcoin offered an innovative option to citizens disenchanted with the existing monetary system. The relatively new digital currency offered not only decentralization but also a limited money supply — all working within an anonymous peer-to-peer, ledger-based transaction system. As such, it skirted concerns around the over-printing of money, privacy, and efficiency. It also implies that, perhaps, a government is not needed for the currency system. However, bitcoin has a lot of issues that come in the way of it becoming a mainstream currency. As of July 17, 2017, the bitcoin blockchain size was 125 gigabytes, which cannot be conveniently stored in a single mobile phone. To get around that these ledgers get stored only on few exchanges thus belying the promise of complete disintermediation. Bitcoin fails to perform all the three functions of a currency in its entirety. Although bitcoin meets the criteria as a medium of exchange, it fails as a store of value and a unit of account. Further bitcoin is built with scarcity in mind and leads to deflation. Some level of inflation is required to stimulate investment"

Another blockchain-based project, Ethereum [30] makes use of Ether as currency which makes use of "smart contracts" to set rules/checks for transfers.

The emergence of Cryptocurrency as a challenger to Fiat Currencies forced Governments to explore the possibility of issuing digital currencies via their central banks. There are several studies on motivations and mechanisms for Central Bank Digital Currency (CBDC) in recent years [31-34]. CBDC, like physical currency, could be made widely available to firms and households. Like reserves, CBC could potentially earn interest. Kessler and Sanches [31] consider the question whether the Central Banks should issue digital currencies. In their model, they look at two scenarios: (i) Digital Currency as a replacement for physical currency and (ii) Digital Currency in the form of bank deposit and study the macro-economic effects of both these designs. They conclude that "the introduction of a central bank digital currency would represent a potentially historic innovation in monetary policy. If households and firms choose to hold and use significant quantities of such a currency, it could lead to a substantial shift in aggregate liquidity, that is, in the types of assets that are used in exchange and that carry a liquidity premium. While there has been much recent discussion of this issue in policy circles, the macroeconomic implications of this shift are not well understood".

The IMF Staff Design Note of November 2018[32] analyzes the topic of CBDC in great detail. They look at both the scenarios i.e., CBDC being token-based or account-based. They consider the ability to make the anonymous transaction as an important criterion that could attract users. They emphasize that "the role of Central Bank in CBDC continues to be (i) As a *unit of account,* money is an important public good that requires price stability in all economic circumstances. (ii) As a *means of payment,* money must be universally available and verifiable as well as efficient, while

Page **7** of **37**

ensuring appropriate consumer protection and minimal cost to taxpayers. (iii) As a *store of value,* money must be as secure as possible, but it must also allow for efficient allocation of resources. In addition, central banks will prefer forms of money that support, or at least do not undermine, three other public policy goals: financial integrity, financial stability, and monetary policy effectiveness. They say that CBDC should compete with cash, commercial bank deposits, narrow finance, and cryptocurrencies."

The IMF Staff Design Note further states that "Narrow finance solutions cover various new forms of private money backed one for one by central bank liabilities, either cash or reserves**.** These offer stable nominal value, security, liquidity, and potentially close to a risk-free rate of return. The parallel here is with currency boards (such as in Hong Kong SAR) or metal-backed banknote systems (such as the gold standard). Two versions of narrow finance solutions are relevant. The first is *stored value facilities17* such as AliPay and WePay in China, Paytm in India, M-Pesa in Kenya, and Bitt.com in the Caribbean. These provide *private e-money* to users against funds received and placed in custodian accounts. Transactions occur between electronic wallets installed on handheld devices, can be of any size (although they are usually not large), and are centrally cleared, but are restricted to participants in the same network. However, holding these forms of money entails some risk. Nonetheless, this segment is gaining widespread and very rapid acceptance. The second version of narrow finance solutions—*narrow banks*—is only beginning to materialize. It covers institutions that invest client funds only in highly liquid and safe government assets—such as excess reserves at the central bank—and do not lend. However, they allow payments in their liabilities through debit cards or privately issued digital money". Overall, the note finds no universal case for CBDC adoption as yet.

Bank of Canada Staff Discussion Paper from November 2017[33] analyzes the motivation for introducing CBDC. They say "in light of recent technological developments a central bank could consider providing digital currency to the public through centralized accounts on its books. Conceptually, this would extend the provision of reserves, currently accessible only to certain financial institutions, to the general public. In this case, the central bank can be seen as a "narrow bank" providing accounts to the general public and allowing account holders to use the balances in these accounts to make payments over the central bank's ledger. Alternatively, a central bank could issue a digital currency in a decentralized manner, similar to how physical cash is distributed".

In their paper, they list the motivations of issuing a CBDC as "(i) Ensure adequate central bank money for the public and preserve central bank seigniorage revenue (face-value of currency – cost to produce it). (ii) Reduce the lower bound on interest rates, and support unconventional monetary policy. (iii) Reduce aggregate risk and improve financial stability. (iv) Increase contestability in payments. (v) Promote financial inclusion. (vi) Inhibit criminal activity". Among these improving financial stability and increasing contestability in payments seems most likely to provide a sound motivation to issue CBDC in Canada. About others, they conclude that there are alternative mechanisms that fulfil these goals equally well as CBDC. They consider promoting financial inclusion could be an important consideration in emerging economies but does not provide motivation in the Canadian case. In our opinion, considering the alternatives available, improving financial stability can be a unique value that is provided by CBDC for all countries.

Bordo and Levin in their National Bureau of Economic Working Paper [34] contend that "CBDC can serve as a practically costless medium of exchange, secure store of value, and a stable unit of account. To achieve these criteria, CBDC would be account-based and interest-bearing, and the monetary policy framework would foster true price stability."

The research reviewed leaves opportunities to explore the design of a digital currency without building a public blockchain as blockchain-based implementations suffer from several limitations. In particular Leger-less solutions with decentralized architecture are worth exploring.

Further, if we have payment banks (narrow finance institutions) manage digital currencies then they are not vulnerable to defaults the way the commercial banks that are into lending. A digital currency issued by Payment Banks can address the motivations identified with CBDC such as Financial Stability, contestability, financial inclusion, and prevention of illicit transactions equally well. India already has the policy framework for setting up more payment banks which can be further strengthened.

## 3. Decentralized Digital Currency System (DDCS)

The classical Blockchain-based cryptocurrencies suffer from the following limitations:

- There is a single all-encompassing ledger. Having copies of everything everywhere is not a good way to design distributed solutions. In the event of any security breach however unlikely it is, the impact cannot be contained to a limited set of users.
- Secondly, as the ledger grows, performance suffers and may lead to scalability issues
- It takes minutes to confirm and add a new block. Compared to this Visa Networks perform thousands of transactions in seconds.
- Even though third parties are not there in a technical sense, without miners the network will not work.
- Without a role for Government/regulators, it is hard to convince people to adopt crypto-currencies.
- Recovery is harder when things fail. People have no one to turn to in case of disputes.

Further, the problems associated with power consumption to enable mining and use for illegal activities are too well-known. A huge number of bitcoins are lost due to the lack of a third party who can resolve issues such as the loss of keys and passwords.

Many of the problems cited above stem from the requirement to maintain a public chain of hashed transactions that serves as a ledger. In this paper we explore designing a digital currency without making use of ledger/blockchain but continuing to use the Merkle Trees to encapsulate transactions. There are already digital currency solutions that are based on cryptography but do not use blockchain such as Ripple [35]. Ripple stores transactions in the form of a hash tree and use multiple validation servers to validate transactions. Ripple is primarily to be used for inter-bank transfers.

## 3.1 Preliminaries

A hash function takes an arbitrary length string as input produces a fixed-size output (e.g., 256 bits). An important property of the hash function is collision resistance. It is computationally not feasible to find x and y such that x != y and H(x) = H(y). However, for a weak hash function, a collision may be possible (MD5, SHA1). This property enables a sender to send hash as a message digest. If we know H(x) = H(y), it is safe to assume that x = y. The other property Hash Function have is hiding i.e. given h(x) it is impossible to find x. This property helps in an application called commitment. Here a value is sealed in an envelope as a commitment and then revealed letter. Anyone can validate that what was revealed was the same as what was committed. This can help in non-repudiation. The third property of hash functions is puzzle-friendliness which is well-exploited in bitcoin currencies.

Merkle Tree is a binary tree that makes use of Hash Functions. Each leaf node holds the hash of a data block. Internal nodes hold the hash of the concatenated hashes of their children. Every node hashes the nodes below it and the hashes are computed upwards till a root-hash is generated. The root hash is signed. This scheme is based on the assumption that a safe/trusted way exists to share the root of the tree between the signer and the verifier. The advantages of Merkle Tree are that it can hold many items and one need to remember just the root hash. It can verify whether a block belongs to a tree in O (log n) operations, where n is the height of the hash tree. To verify the integrity of any data block, the whole tree of hashes does not need to be transmitted to the verifier. A signer transmits the hashes of only those nodes which are involved in the authentication path of the data block under consideration. Merkle Tree is illustrated in Figure 1 below. In Figure 1 data blocks are represented by the gray, rounded rectangles and hash blocks in light gray squares. The arrows indicate taking the hash after concatenation. The root (dark blue) is a signed hash.

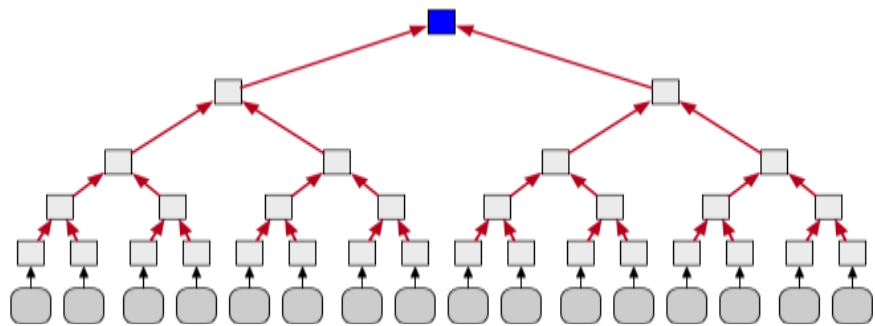

**Figure 1: A Merkle tree**

The Merkle Tree was introduced in a seminal paper by Ralph C Merkle in 1979[36] as an improvement on previous research on digital signatures by the likes of Lamport [37] and others [38,39]. Digital Signatures continues to be a fertile area of research. In his recent work, Alzubi [40] made use of Lamport Merkle Digital Signature as an authentication tool in an IoT environment.

B+ trees are a special case of B trees as shown in Figure 2. They are n-ary trees. Internal nodes only hold keys, they do not hold data. Data always stays in the leaf nodes. Leaf nodes are connected through pointers to form a kind of linked list. This linkage helps in the sequential traversal of the data. Besides keys, leaf nodes also hold the hashes of the data records pointed by corresponding keys. An example for B+ Tree is depicted in Figure 2.

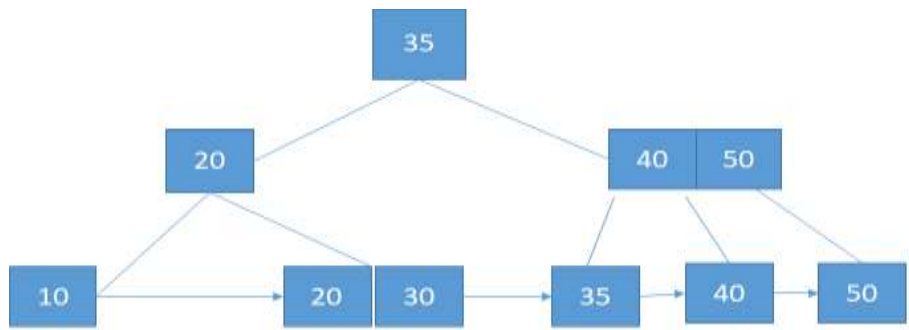

**Figure 2: B+ Tree example**

MHT (Merkle Hash Tree) is designed by replacing binary trees with B+ trees in the original Merkle's Signature Scheme. MHT is explained at a later stage in the paper.

## 3.2 Merkle Tree as Authenticated Data Structure

Authenticated Data Structure (ADS) is a model of computation where an untrusted responder answer queries on a data structure on behalf of a trusted source and provides proof of the validity of the answer to the user. Tamassia [41] discusses the use of Authenticated Data Structures in Data Replication Scenarios. This is before the emergence of Cloud Computing and Data Outsourcing as a common phenomenon.

Another classical use case for ADS is Data Publishing. Here the owner delegates the role of satisfying user queries to a third-party publisher. As the publisher may be untrusted or susceptible to attacks, it could produce incorrect query results. Pang et al. [42] introduce a scheme for users to verify that their query results are complete (i.e., no qualifying tuples are omitted) and authentic (i.e., all the result values originated from the owner). The scheme supports range selection on key and non-key attributes, project as well as join queries on relational databases. Nuckolls et al. [43] propose a general model for authenticated data structures by taking the case of online publishing. Here Authentic publication allows *untrusted* publishers to answer queries from clients on behalf of trusted offline data owners in a secure manner. Publishers validate answers using hard-to-forge *verification objects* (*VO*s), To make authentic publication attractive, The VOs need to be small, efficient to compute, and efficient to verify. They make a study of the suitability of Directed Acyclic Graphs as VOs. Figure 3 below [43] illustrates the authentic publication scheme.

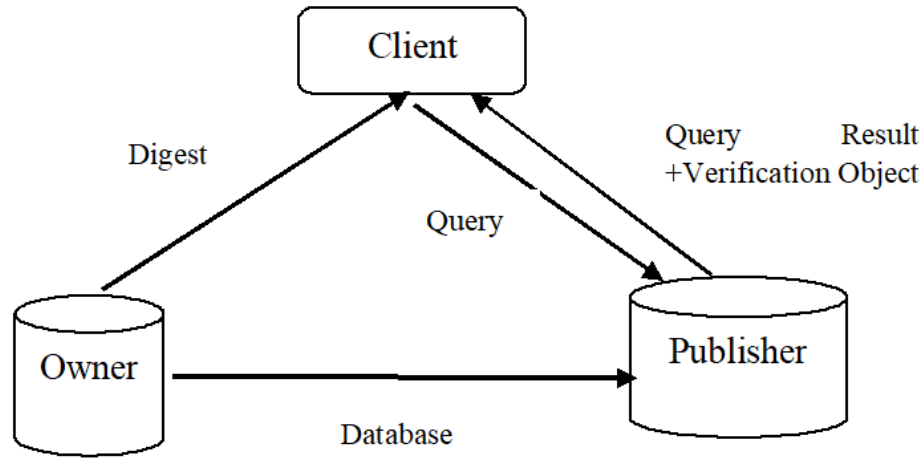

**Figure 3: Authentic Publication Scheme**

Sion [44] discusses mechanisms for query execution assurance in Outsourced Database (ODB)'s. They imagine scenarios a service provider may avoid executing queries that are compute-intensive need a lot of storage and return inexact responses and in other cases service provider does not return the results even when he has them. These two pertain to Completeness and Correctness. Feifei Li et al. [45] study Outsourced Database Systems and define a variety of cost metrics associated with them. They focus on scenarios wherein data owners periodically update the data residing in the servers. In the process, they explore query freshness which was not explored in earlier literature. Thus, conceptually Data Integrity evolves as a triad consisting of correctness, completeness, and freshness. According to them, "correctness means that the client must be able to validate that the returned records do exist in the Data Owner's database and have not been modified in any way by DSP. Completeness means that no answers have been omitted from the result by DSP. Finally, freshness means that the results are based on the most current version of the database, that incorporates the latest updates by Data Owners".

Xie and Wang [46] address the problem of query integrity. Generally auditing the query results means the clients have to manage the data locally or database engine to be modified to generate authenticated results. In their work, they insert a small number of records into an outsourced database, using both randomized and deterministic approaches so that the integrity of the system can be effectively audited by analyzing the inserted records in the query results... Furthermore, they show that their method is provable secure, which means it can withstand any attacks by an adversary whose computation power is bounded. In the follow-on work [47] they discuss mechanisms of providing freshness guarantees of outsourced data.

When records are inserted into a database with deterministic approach, a Verification Object (VO) can be generated, but it requires modification to the database. The VO is returned along with query results to clients that enable them to verify the integrity of query results. Third-party service providers are reluctant to do such modifications. This is corroborated by Pizzette and Cabot [48], who did a market assessment of the Database as a Service Provider model. They concede that no existing cloud database services support integrity checking. The probabilistic approaches do not require modifications to the database engine but the query integrity provided by them is on the lower side.

Wei Xin and Ting Yu [49] present an efficient and practical integrity assurance scheme without requiring any modification to the DBMS at the server-side. They develop novel schemes to serialize Merkle B-tree-based authentication structures into a relational database that allows efficient data retrieval for integrity verification.

The next scenario for the application of ADS is in Data/Databased Outsourcing. Here Data Owner (DO)s outsource data management to a Data Service Provider (DSP) who respond to queries of Data Clients. DSP hence is also called a server. DSP has unlimited access to the data to make it possible for the DSP to forge the data. Thus, the data-owners and clients need a mechanism to authenticate the integrity of data managed by Data Service Providers/Serers. While designing the outsourcing technique the following overheads become critical: Computation Overhead for the DO, Computation Overhead of the DSP, Storage Overhead of DSP, Computation overhead of the client, and Storage overhead of the client. The Database outsourcing model proposed by Wei Wei and Ting Yu is shown in Figure 4 below. Other models may differ as far as the flow and location of authentication data. Generally, Data Clients may submit only queries do the verification locally once they receive the verification object. See Figure 4 below.

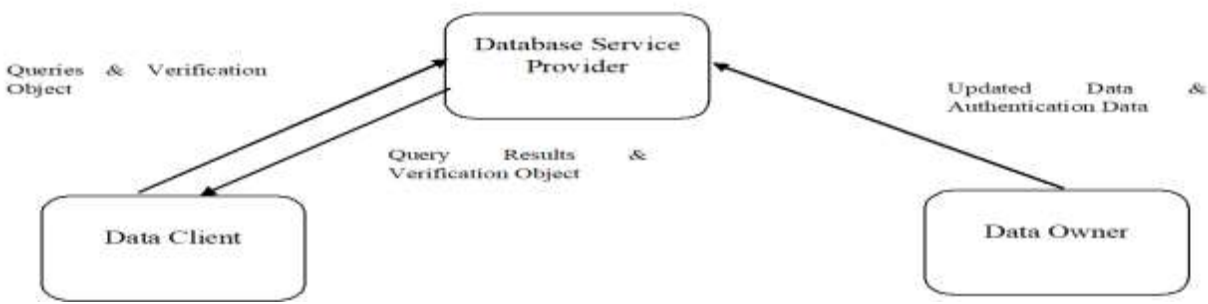

**Figure 4: Data Outsourcing Model**

Niaz and Saake [50] discuss the issue of data integrity in a data outsourcing environment. According to them, the simplest data integrity technique could be to individually sign all the tuples in a data table and send the data to clients with the signatures. Clients can then check the integrity of the data by verifying the signature of the DO for the associated tuple. However, DSP can delete some valid tuples from the data or send incomplete data-set to the client and the client would never be able to establish this fact. Some of the authentication schemes they refer to include ssignature aggregation which requires modifying signatures of all the records, the authenticated skip lists, and a modified Merkle Hash Tree (MHT) based scheme where the main MHT is divided into smaller MHTs and the root hashes of these sub-trees are signed. The purpose of the division in smaller MHTs is to avoid the unnecessary calculation up to the root of the main hash tree.

Niaz and Saake [50] make use of MHT based data integrity techniques which are based on two sub-components, i.e., Merkle's Signature Scheme and B+ Trees. The focus of their paper is how to store the MHTs in database tables efficiently. For that, they propose a radix path identifier to be associated with each node. Niaz and Saake propose two ways of storing MHTs in Databases: (i) Single Authentication Table and (ii) Level-based Authentication Table and then conclude that the latter is superior. To enable this, they come up with a radix-path identifier scheme.

Dongyoung Koo et al. [51] discuss Merkle Tree-based authentication. In a typical Merkle-Tree-based authentication, there are two entities, prover[P] and verifier{V}.
- Prover P is an entity that attempts to convince the other party (i.e., the verifier V) that it owns all of the data. To conserve network bandwidth, the prover sends a small piece of verifiable information instead of all of the content.
- Verifier V is another entity that tries to determine whether prover P's claim is correct or not. To reduce storage requirements, the verifier usually stores only the value of the root node of the Merkle tree instead of all nodes of the tree.

Notably, the Merkle tree-based online authentication is a protocol that verifies that the prover and verifier own the same data. Unlike public verification, therefore, it assumes that the verifier has some secret (i.e., not publicly available) information about the data to be validated. In the above scheme, the verifier V only stores the value of the root node of the tree and removes the rest of the metadata once the tree is constructed. On the other hand, the prover P is required to generate a series of (different) hash values leading to a value of the root node that is identical to the one held by the verifier with each authentication cycle. Further, Dongyoung Koo et al. [51] analyse threats due to adversaries eavesdropping on the communication between Prover and Verifier and discuss a mechanism to prevent information leakage.

Wang et al. [52] make use of Merkle Tree verifications to ensure the integrity of Map Reduce computations in a Hybrid Cloud environment. Here the "prover is the public cloud and the checker is the private cloud. The verification is performed in a commit-and-verify manner. We break the entire verification into four steps, i.e., the commit step, the challenge step, the prove step and the verify step. In the commit step, the prover constructs the Merkle Tree and creates a root-hash, Then the prover sends the root hash to the checker as a commitment. (In the case of DDCS the peers will act as checkers to each other). In the challenge step, the checker provides the information on sample nodes and asks for complementary nodes information from prover. In the proving step, upon receiving the challenge, the prover sends the complementary nodes information. In the verify step, the checker recomputes the root value."

Paris and Schwartz [53] in their 2020 paper, propose Merkle Grids as a novel data organization that replicates the functionality of Merkle Trees while reducing transmission and storage costs by up to 50 percent. All Merkle grids organize the objects whose conformity they monitor in a square array. They add the row and column hashes to it such that (a) all row hashes contain the hash of the concatenation of the hashes of all the objects in their respective row and (b) all column hashes contain the hash of the concatenation of the hashes of all the objects in their respective column. In addition, a single signed master hash contains the hash of the concatenation of all row and column hashes. They also proposed Extended Merkle grids with two auxiliary Merkle trees to speed up searches among both row hashes and column hashes. While both basic and extended Merkle grids perform authentication of all blocks better than Merkle trees, only extended Merkle grids can locate individual non-conforming objects or authenticate a single non-conforming object as fast as Merkle trees. Merkle Grid can be useful to store system states at different points in time.

Fang, Laio, and Lai [54] discuss an efficient share verification method using Merkle trees. Here the root and authentication paths of a Merkel tree are used to verify shares between the participants so that they can reconstruct secrets correctly after verifying and eliminating the fake shares. This approach is an improvement on Shamir's secret sharing scheme [55].

In recent work, Xu et al. [56] make use of privacy-preserving streaming authenticated data structures in a cloud environment for the healthcare cyber-physical system. In their work, they emphasize the importance of simultaneously paying attention to privacy and integrity.

Etemad et al. [57] propose the use of hierarchical authenticated data structures and claim them to be superior to the non-hierarchical structures.

Off-the-shelf commercial databases do not support storing of authenticated data structures. Penino et al. [58] propose the use of overlay indexes to address this lacuna and describe a structure called DB-Tree to realize the overlay index.

## 3.3 Decentralized Digital Currency System Proposal

We propose a novel Decentralized Digital Currency System (DDCS) which provides an alternative approach to blockchain-based currencies, without requiring mining yet taking care of security consideration. DDCS consists of Currency Clients, Currency Manager, and Integrity Manager. In addition, DDCS supports Data Clients, who can only query and interrogate the transactions in DDCS, provided they are legally authorized. Here Currency Clients are transacting parties (individuals and businesses). The Currency Manager role can be done by a Payment Bank which can deal with the digital δ-Money. Integrity Manager is a role integral to the DDCS platform that cross-validates transactions. Its role can be considered as that of a verifier which operates at the system level. The role of Integrity Manager can be done by DDCS operator as a native service that is extended to multiple Payment Banks.

We can compare DDCS with Data Publication and Data Outsourcing Scenarios discussed in 3.2. In both Data Publication and Data Outsourcing, the Data Management is outsourced to a third party i.e., Data Publisher or Database Service Provider respectively. Here the third party is untrusted. In the case of DDCS, the parties (Currency Clients) themselves may be untrusted by other Clients as well as the Currency Manager. They need assurance by the Currency Manager as far as the currencies they are dealing with and Integrity Manager as far as the transactions they have entered into. This needs to happen while maintaining a disintermediated operation in a decentralized system. When it comes to a distributed system with a huge number of participants operating using say a mobile application, trust can be broken down into intent and competence. There may be some clients who have malicious intentions and attempt to gain by acting in an adversarial manner. The other aspect is competence where the parties need the ability to protect the integrity of data they are managing; this needs to be enabled by the system with appropriate authenticated data structures. As laid out in the last section, data integrity should cover completeness, correctness, and freshness. The Data Clients who may take decisions based on this data have a particular stake in this.

Some of the important guiding principles to arrive at the architecture of DDCS are described in Table 1 below.

### Table 1: Architectural Principles of DDCS

| Sr. No | Principle | Description |
|---|---|---|
| 1 | Separation of Concerns | The Currency Manager's concern is currency conservation. No invalid currency should circulate. All the circulating currency is accounted for among the Currency Clients at any point in time. To record any change in balance, The Currency Manager needs the balance provenance information. <br><br> Integrity Manager's concern is to establish a consistent account of peer-to-peer transactions among peers and thereby at the system level. To validate any peer-to-peer transaction Integrity Manager needs transaction provenance information. <br><br> Currency Clients have a stake to ensure that the transactions and balances are correctly recorded. <br><br> Data Clients have a stake to ensure that the results of their query are complete, correct, and fresh and they get to know if their funds are mis-utilized in a timely manner. |

| 2 | Distributed State Management | Currency Manager maintains the state related to currency balances and state-related Currency Clients' enrollment and authentication. |
| --- | --- | --- |
| | | Integrity Manager retains state related to transactions to continually validate the transactions. |
| | | Currency clients maintain with them the state on transactions engaged by them with peers and the resulting balance changes. |
| 3 | Peer-to-Peer Architecture | All Currency Clients engage as peers whenever they engage in transactions. They register themselves as peers with mutual consent recorded with the Currency Manager. Currency Manager also acts as a peer as and when it engages in a transaction with clients. Even though most transactions may happen between two peers, the architecture is extensible to transactions involving multiple peers. |
| 4 | Optimistic Commit | The peers commit the transactions to each other without waiting for the Integrity Manager. It stays committed unless an exception is raised by the Integrity Manager. |
| 5 | Disintermediation | The peers can transact without involving any other intermediary during the transaction. The ability to transact remains as long as peers can connect mutually. |
| 6 | Eventual Consistency | When there is a huge number of peer-to-peer transactions, the system may reach a consistent state eventually. |
| 6 | Recoverability | When any one of the peers loses the state information it can be reconstructed using the state information with other peers. Integrity Manager and Currency Manager can reconcile their states in case of any anomalies. |
| 7 | Reparation | In the event there is a malicious client, it may run afoul of currency conservation and will get suspended. Any illegitimate transaction can be reversed by readjusting the balances. |
| 8 | Privacy | The visibility of information can be configured. Information sharing can be limited to only what is required to establish provenance. Any information which is shared between software agents is not necessarily available to the human operator such as peer balance information. |
| 9 | Anonymity | Anonymity or pseudonymity is not a goal. However, private information is not shared unless legally required. |
| 10 | Security | The Integrity Manager and Currency Manager can be housed in two different institutions i.e., DDCS operator and Payment Bank respectively. The system, in general, should be robust to prevent cyber-security attacks on clients, managers, and the state information maintained by them. |

Based on the architecture principles laid out above, the roles and responsibilities of DDCS participants are described as follows:

- The Currency Clients are the prime users of DDCS. They transact with other peers. They get their δ-Money from the Currency Manager.
- The Currency Clients are expected to register as peers with Currency Manager as a pre-condition to transact with each other.
- The Currency Clients are expected to report their balances to the Currency Manager after transactions, along with a Verifiable Object (VO).
- The Currency Clients are expected to report transactions along with a Verifiable Object (VO) to the Integrity Manager.
- The Currency Manager enrolls the Currency Clients as participants in DDCS and provides them a unique Client Id. The Currency Manager also registers them as peers with mutual consent.
- The Currency Manager issues δ-Money to Currency Clients in exchange for some form of deposit/item/instrument/money substitute that has a particular valuation. Currency Manager accepts any unused currency from Currency Clients to be swapped for equivalent value.
- The Currency Manager authenticates the peer client to other peers either at first interaction or whenever they suspect any anomaly.
- The Currency Clients can be divided into domains and zones by Currency Manager.
- A transaction amount limit can be associated with each client by the Currency Manager.
- Currency Manager is also authorized to disenroll/suspend Currency Clients after following due process, under conditions laid out during enrollment.
- The Currency manager is responsible for Currency Conservation across DDCS. It does that by tracking of δ-Money balances with Currency Clients at any given point in time.
- The Integrity manager is responsible to cross-validate every transaction between peers and in the process validate all the transactions in the system.
- In the event of any issue during validation, Currency Manager and Integrity Managers raise alerts taking the system to the reparation phase in part or full.
- The Data Clients may be Corporates, Banks, Government Departments, and Public Institutions who may want to ascertain the authenticity of transactions and balances reported to them by the Currency Clients when they have the authorization and need to do so.

Figures 5 & 6 below, illustrate the workflow and dataflow among the participants in the operation of DDCS respectively. Figure 7 illustrates Transaction and Balance Authentication by Data Clients. In Figure 7, the Data Client role may be played by a Government Department or a Company where an employee is operating an expense report on their behalf. Here only authorized vendors or beneficiaries are added as Peers/Counter-parties. The role can also be played by a Bank which credits the loan amount to a beneficiary and has a stake in knowing that it is utilized only for the intended purpose and in the pace and manner expected by the lender.

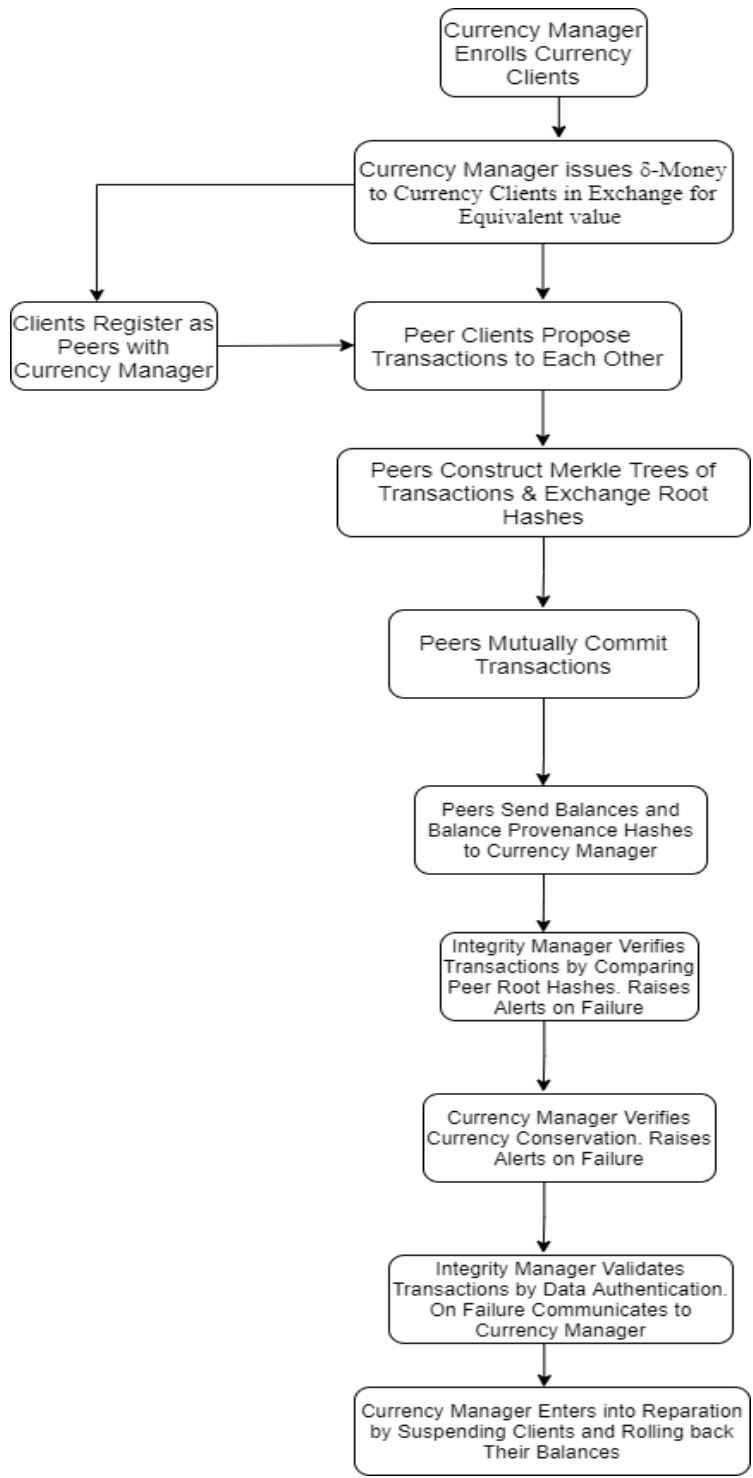

**Figure 5: DDCS Workflow**

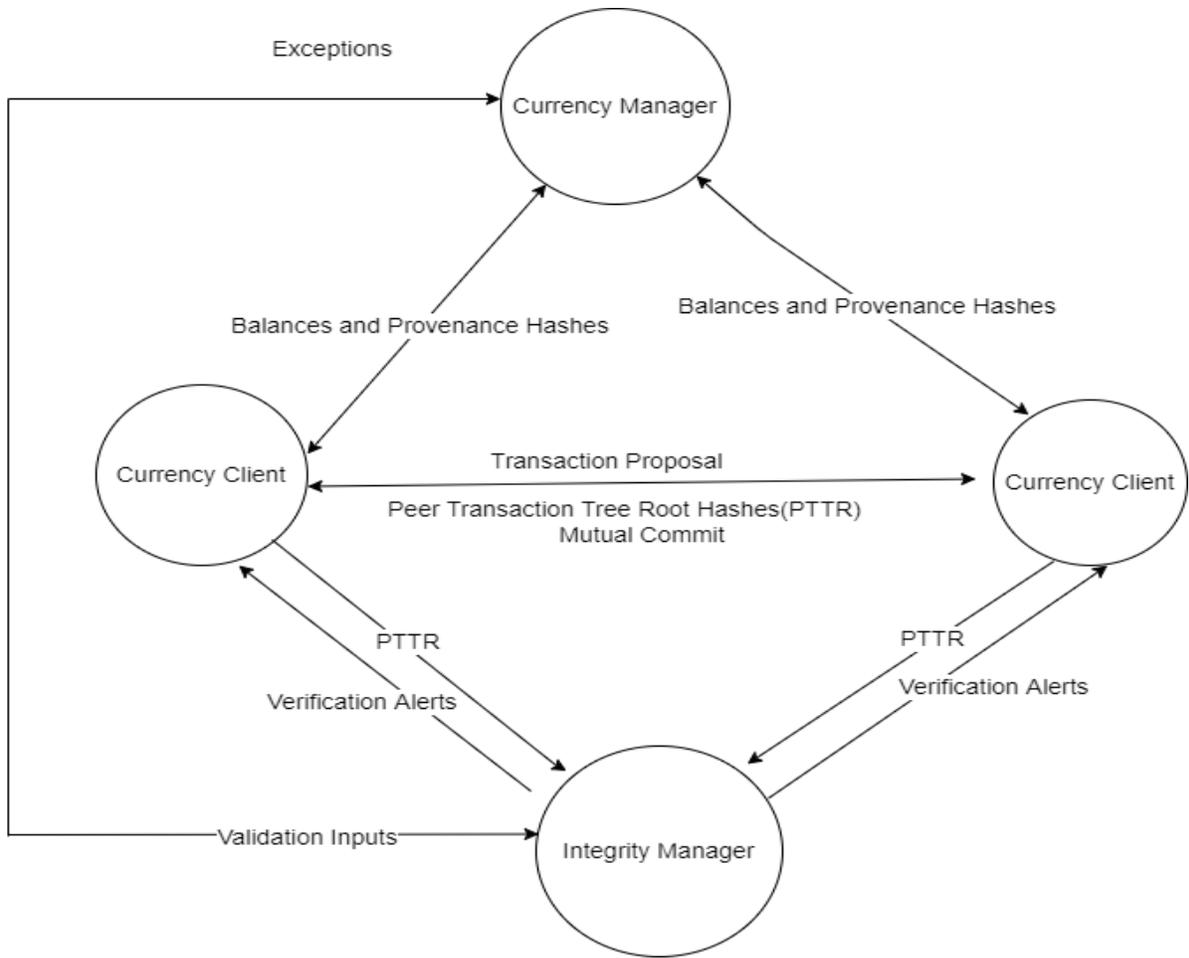

Figure 6: DDCS Dataflow

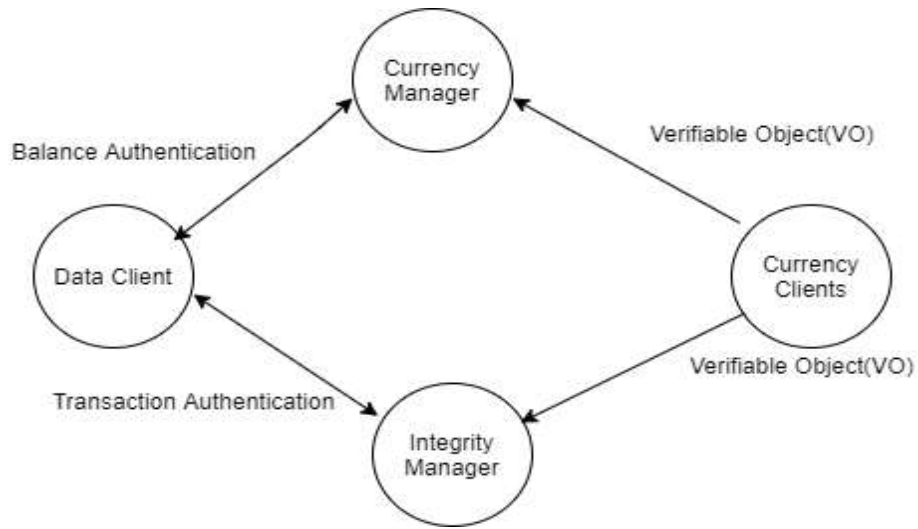

Figure 7: Data Client Operations

The DDCS operations are detailed as follows:

1. Each Currency Client transacts with peer clients as follows:
    - Clients need to register themselves as Peers with mutual consent and confirmation.
    - Peers propose transactions mutually that have credit/debit amount and balance amounts as well as balance-provenance information.
    - They construct identical Peer-to-peer transaction trees and exchange root hashes.
    - The transaction is committed only if both the peers can generate the same root hash.
    - Peer-to-peer transaction state is maintained by each peer in a Merkle Tree as explained in the next step.
    - Clients should securely keep the root hashes, accessible only to corresponding peers.
2. Each Currency client stores a Merkle Tree for every peer client it transacts with. Every Transaction of currency transfer is expressed as a credit-debit transaction pair. The transaction stored not only contains the amount transferred but also the balance amounts at the time of effecting the transaction as well as Balance Provenance Hashes. The transactions are hashed and stored in a Merkle Tree. Each tree can thus generate net credit or debit amount relative to the peer at any point in time. Table 2 and 3 below record transactions at the peers and the corresponding Merkle Tree is represented in Figure 2 below. The balance hash management is explained further below.

### Table 2: Transaction Table of Alice

| TransId | Timestamp | Currency Client Balance | Client Balance Hash | Peer Client(s) | Peer Client(s) Balance Hash(es) | Debit/Credit Amount | Currency Client New Balance |
|---|---|---|---|---|---|---|---|
| 10.1 | | | | Bob | | | |

### Table 3: Transaction Table of Bob

| TransId | Timestamp | Currency Client Balance | Client Balance e Hash | Peer Client(s) | Peer Client(s) Balance Hash(es) | Debit/Credit Amount | Currency Client New Balance |
|---|---|---|---|---|---|---|---|
| 10.2 | | | | Alice | | | |

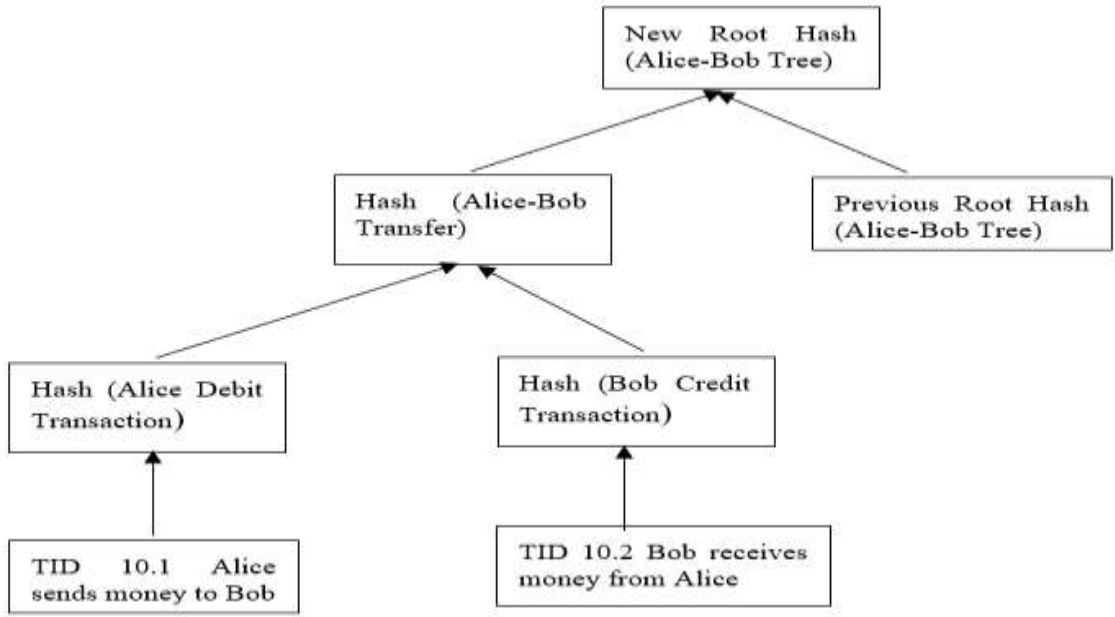

**Figure 8: Peer-to-peer Transaction Merkle Tree**

3. Each Currency client stores all its transaction balances as s a tagged Merkle Hash Tree (MHT) [23]. Figure 9 depicts Client Balance Merkle Hash Tree. In Figure 9 the transactions that record balance changes and peer-client ID are stored in the leaf node and hashed upwards to generate a parallel Merkle Tree of balances generating its root-hash that is associated with Currency Client balance at any point in time. Here each node has the key, pointers to other nodes as well hash of transactions located below the root, sub-root, or leaf as applicable. Each Leaf node is tagged with the root-hash of Peer Transaction Tree that was responsible for balance change.

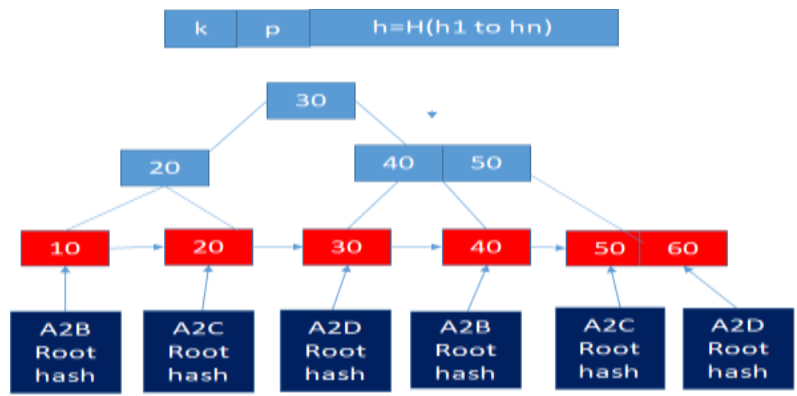

**Figure 9: Client Balance Merkle Hash Tree**

In Figure 9 above, the Merkle Hash Tree depicted is nothing but a combination of Merkle Tree and B+ Tree. B+ Tree is a B-Tree where Leaf Nodes can be chained, whose traversal can give all details in an expedited manner. B-Tree structure enables efficient storage of voluminous transaction information as well as query performance.

4. Currency clients report the root hash of peer transactions to Integrity Manager, which is supposed to verify that both the peers return the same root hash. If any issues Integrity Manager raises an alert on verification failure. In the case of multiple peers, this may have added complexity but that can be tackled.
5. Currency clients report the current balance along with the root hash to the Currency Manager. They also need to report the balance provenance hash of the peer client. The Currency Manager should verify Currency conservation. The currency manager can make use of suitable mechanisms to track balances and securely store them. Temporal Databases [59] can be an attractive option to keep track of balances and validate conservation in case of voluminous and concurrent transactions. See Table 4 below. This kind of tracking can be useful to support audits as well as reconfirming balance conservation in any time window in the past.

**Table 4: Temporal Database of Client Balances**

| Transaction Time-stamp | Client ID | Valid Balance | Valid From | Valid To | Remarks |
|---|---|---|---|---|---|
| T1 | 1 | 1000 | May 17 2PM | ∞ | Initial Balance |
| T2 | 2 | 2000 | May 17 2 PM | ∞ | Initial Balance |
| T3 | 1 | 1000 | May 17 2PM | May 18 3 PM | Updated Record |
| T3 | 2 | 2000 | May 17 2PM | May 18 3 PM | Updated Record |
| T4 | 1 | 1500 | May 18 3 PM | ∞ | Updated Balance |
| T4 | 2 | 1500 | May 18 3 PM | ∞ | Updated Balance |

6. Integrity Manager cross-validates balance information between Currency Manager and Currency Client, if any issues it raises an alert indicating validation failure.
7. Currency Clients can return their currencies in part or full to Currency Manager, which takes the form of yet another peer-to-peer transaction.
8. The following modalities are proposed to institute recovery in the event of failures:
    o If any client fails, its transactions can be recreated by recovering all its partner hashes.
    o If any client fails, its balance can be recovered if the balances of all others are known and the total currency in the system is known. Its latest balance can be known from the Currency Manager.

- If Currency Manager fails, the currency in the system can be recomputed by knowing the balances of clients subject to Integrity Manager validation.
- If Integrity Manager fails a new instance can be created.
9. Following refinements can be considered depending on the operating conditions.
    - Each Currency Client can store the B-Tree of Balances on a separate card/media or any persistent store, which can be used to redeem the currency from Currency Manager, in the event client software is corrupted.
    - The Peer-Transaction-Trees can be reset and new trees can be initiated. The Peer-Transaction-Trees can be archived in parts by keeping only the recent transactions in memory.
10. Integrity Manager and Balance Manager can capture the system state as Merkle Hash Grid during the quiet period as shown in Table 4 below. Here the cells on the diagonal are the hashes that of the Merkle Hash Tree of Currency Clients. Then the cells other than diagonal contain Peer Transaction Tree Root (PTTR). The PTTR Hashes are recorded only in the first Peer's row. For every row, hashes are concatenated and a row-hash is created. Similarly, column-hashes are created by concatenation. The grid hash is created by concatenating the concatenated row hash and the concatenated column Hash. all rows hash and all columns Hash. The Merkle Hash Grid is due to Pâris and Schwarz [51]. The way we have designed the Merkle Hash Grid is different from the way they in the sense that there are no data blocks here. Rather we are using the Root Hashes of Trees that represent the system state directly. The Grid Hash can be signed by the Integrity Manager. See Table 5 below.

**Table 5: Merkle Hash Grid for DDCS**

| System State – 3 Peers Transacting in DDCS | | | | |
|---|---|---|---|---|
| | Peer1 | Peer 2 | Peer 3 | Concatenated Row Hash |
| Peer 1 | MHTR 1 Hash | PTTR Hash (1,2) | PTTR Hash (1,3) | Row 1 Hash |
| Peer 2 | | MHTR 2 Hash | PTTR Hash (2,3) | Row 2 Hash |
| Peer 3 | | | MHTR 3 Hash | Row 3 Hash |
| | Column 1 Hash | Column 2 Hash | Column 3 Hash | Concatenated Column Hash |

Grid Hash

In summary, the proposed DDCS can store transactions in a tamper-proof manner. The DDCS system can help track the distribution of currency among the participants as well as their transactions. The system state can be captured periodically using Merkle Hash Grid. At the same since there is no single all-encompassing ledger, the system can provide better privacy, security, scalability, throughput, and latency. We have also proposed mechanisms to keep track of balance provenance as well as system state. The DDCS can support any plausible need for Governments to interrogate illegal transactions. The privacy needs can be fine-tuned based on the laws of the land.

## 4. Discussions

While designing the outsourcing technique the following overheads become critical: Computation Overhead for the DO, Computation Overhead of the DSP, Storage Overhead of DSP, Computation overhead of the client, and Storage overhead of the client. We can use a similar approach for DDCS. As of now there are no implemented DDCS systems, hence we do qualitative analysis. We compare DDCS with traditional modes of Digital Payment via account-based digital money or through wallets. Here we do not consider communication or computation overheads associated while serving Data Clients. In DDCS, double-spend is not going to go undetected as currency manager tracks conservation of balances continually. See Table 6 below.

**Table 6: Overheads and Benefits in DDCS**

| Overheads | Currency Client | Currency Manager | Integrity Manager | Remarks |
|---|---|---|---|---|
| Computation Overhead | Overhead of Creating Peer Transaction Tree Root Hash and MHT Root Hash | Computation should be done to check Balance Conservation. | Comparing the Root-Hashes. Reconstructing the MHT's in case of anomalies. Generating Merkle Hash Grid at system level periodically. | Computation distributed among participants. |
| Storage Overhead | Storing PTTs and MHTs in memory/ cache or persistent storage/media. Inactive PTT's can be archived. | Balances can be maintained in a snapshot database or temporal database. Client enrolment information needs to be stored. No transaction information needs to be stored. Transactions are | Storing of Peer Transaction Tree Hashes and transaction information as long as required. Storing of System State. Some information can be archived periodically. No balance | State information distributed among participants. |

|  | | not recorded on Bank Ledger. | information needs to be stored. | |
|---|---|---|---|---|
| Communication Overhead | A Proposal Message and Message with Root Hash needs to be sent to Peer.<br><br>A message each needs to be sent to the currency manager and Integrity manager with MHTR and PTTR respectively. | A message with validated MHTR needs to be sent to the Integrity Manager, after the transaction. Alerts need to be raised in case of anomalies and exceptions. Authentication Messages/Health check messages to currency clients | Raising Alerts and Exceptions to Currency Clients and Currency Manager | Communication can be generally asynchronous and the state can be reconciled eventually. Near-field protocols can be used between peers. |
| Benefits | Currency clients can retain the complete state of transactions engaged by it in an authenticated Data structure in self-custody. | Can focus only on conservation of currency without getting into minutiae of transactions. It is like issuing a trackable fiat currency. | Can focus on its role as a Verifier of Transactions and consistency of Transaction Accounts | Overall a decentralized system with a focus on Data Integrity where privacy and security can be carefully traded off. The whole system works without ledgers. |

Currency clients can decide what granularity of details they would like to share and store with the Integrity Manager. DDCS as a platform can evolve to include value-added services such as protecting client state information from cybersecurity threats or theft. Some critical transactions may have a copy with Integrity Manager so that they can be readily authenticated when the need arises. The peer-to-peer model may make it easier to generate reports focused on peers or peer groups and share them with all the peers.

Next, we compare DDCS with Blockchain Technology in particular the seven limitations cited by Swan [27]. Throughput, Latency, Size and Bandwidth, Security, Wasted Resources, Usability, Versioning, Hard Forks, and Multiple Chains. See Table 7 below

**Table 7: Performance Analysis of DDCS from Design Standpoint**

| Sr. No. | Performance Criteria | Analysis of DDCS from Design Standpoint |
|---|---|---|
| 1 | Throughput | During the transaction, any communication happens primarily with peers. Thus, a large number of concurrent transactions are feasible in DDCS which has an eventual consistency model. The computation overhead at the client level is higher but since it is local, we expect it not to have bearing on throughput |
| 2 | Latency | Latency for Transaction commit is expected to be low. Latency can be impacted when there are alerts and anomalies. |
| 3 | Size and Bandwidth | The transaction state is distributed among the clients. The integrity manager and currency manager only store what is required with the ability to archive the older transactions. The system state is captured by a single Merkle Hash Root Grid. |
| 4 | Security | DDCS has to guard against any adversary impersonating or eavesdropping using available cyber-security tools. Client-end security also needs particular attention. It is important to save the client state from the impersonators. Anonymity or pseudonymity is not the goal of DDCS, but the required privacy can be supported. DDCS also strikes a balance between the role of clients and that of third parties by redefining them to mutual benefit. Compared to bitcoin, DDCS keeps track of account holder details, and hence δ-Money is not permanently lost when one loses his private key |
| 5 | Wasted Resources | DDCS does not use Cryptographic Mining. Validation may require the reconstruction of Merkle Trees or authenticating blocks with only the required information. |
| 6 | Usability | This is a huge strength as DDCS can develop client-focused functionalities and create reports and dashboards with greater flexibility, as all of the transaction data is directly accessible to Clients. |
| 7 | Versioning, Hard Forks, Multiple Chains | The system state is stored in a distributed manner with mechanisms to achieve eventual consistency. Thus, this is not an issue for DDCS. DDCS relies on mutual consensus between peers. |

Next, we look at the positioning of δ-Money with other modes of payment popular in India. See Table 8 below

## Table 8: Comparison of δ-Money with other currencies

| Criteria | Physical Currency | CBDC | Inter-bank Transfer through NEFT/RTGS | UPI | Paytm like Wallet Account | Bitcoin | δ-Money |
|---|---|---|---|---|---|---|---|
| Currency used | Fiat currency in the form of Notes and Coins. | Digital Currency issued by Central Bank | Account-based in Commercial Banks | Account-based in Commercial Banks | Account-based with Wallet Service Provider | Token-based | δ-Money account maintained with Payment Banks |
| Valuation | May lose value due to Government policies. | May lose value due to Government policies | May lose value due to Government policies | May lose value due to Government policies | May lose value due to Government policies | The value fluctuates widely generally appreciated | Can vary as issued by Payment Bank |
| Target Users | Those who seek anonymity & convenience | Those who seek anonymity & want to deal online | Formal Payments generally low volume and high value | Informal Payments generally high volume and low value | Informal Payments generally high volume and low value | Those who seek anonymity | Targeted Peer groups with frequent and routine transactions |
| Ideal Usage Mode | Not suited for high-value transactions | No restrictions | Not suited for low-value transactions | No restriction | No restriction | Suited for High value and low volume transactions | Ideal to manage expense accounts, loan accounts, and beneficiary accounts |
| Intermediation | None required | Depends on whether account-based or token-based | Intermediation needed | Intermediation needed | Intermediation needed | Dependency on Miners | Peers can transact directly followed by post-facto verification. |
| Off-line Transactions | Possible | Not Possible | Not Possible | Not Possible | Not Possible | Not possible | Possible with Near-field communication between peers. |
| Supports Non-repudiation | No | Depends on Design | Yes | Yes | Yes | Yes. But no notion of accounts and owners. | Yes |

| | | | | | | | | |
|---|---|---|---|---|---|---|---|---|
| Double-spend | Not possible. | Not possible | Not possible | Not possible | Not possible | Not Possible. But ensured with great cost | Gets detected in conservation or verification phase | |
| Replay Attacks | Not Possible | Not Possible | Not Possible | Not Possible | Not Possible | Not Possible | Gets detected in conservation or verification phase | |
| Reversibility and Reparation | Not possible | Not possible | Not Possible | Not Possible | Generally, Not Possible | Will lead to a fork of the chain | Possible | |
| Fake/Counterfeit Currency | Possible | Can be prevented | Not Applicable | Not Applicable | Not Applicable | Gets detected before confirmation | Gets detected in conservation or verification phase | |
| Supports Interface to Data Clients | Not Applicable | Not Applicable | No | No | No | No | Yes | |
| Risk & Loss Due to technical impairment | No. Only Due to theft and fake currencies | Yes, if anonymity is desired | Technical impairment addressed with help of bank | Technical impairment addressed with help of bank | Technical impairment addressed with help of bank | Yes. If private keys are lost | Technical impairment addressed with help of Currency Manger. | |
| Risk and Loss due to financial mismanagement | No | No | Yes | Yes | No | No | No | |
| Criminal activities encouraged | Yes | Yes | No | No | No | Yes | No | |

Further, we envisage that Payment Banks can act as Currency Managers and Integrity Manager role is done by DDCS operator. In this scheme, no single participant has all the power. In DDCS, banks manage only currencies and balances. They do not need to record transactions on their ledger. The only entries they record in their ledger are currencies issued and returned to the clients. This is particularly useful when the transactions are high volume and low value. Minimally banks have to manage the conservation of balances using snapshot databases. We have also suggested the use of temporal databases to support the peak load of concurrent and related transactions. The DDCS operator by assuming the role of integrity manager can verify transactions and store them (only to the extent required). Our solution is ledger-less.

Payment Banks can develop their unique revenue models to recover their running costs and make the operation profitable. As the monies held by Payment Banks are not available for lending the possibility of defaults due to lending is not there. The Payment Banks have an option to tie the valuation to the fiat money or a stable valuation by keeping some commodity such as gold or copper as standard or exchange standard. Even when the value is the same as fiat money, using δ-Money saves the cost of printing physical currencies. When Payment Banks are allowed to have a different valuation for δ-Money we can have some kind of sand-box where the wage-earners and pensioners can find a degree of stability for their money. This mechanism in turn can free up commercial banks to focus primarily on lending. This can prove to be a win-win for the Government which has to cope with inflation and demand to grow the economy which at times are contradictory.

δ-Money can be more trustworthy compared to money deposited with a wallet provider as the state of monetary transactions engaged by a Currency Client is entirely maintained in an authenticated data structure with itself. The currency manager is a supervised commercial bank. Integrity Manager is likely to be a Government-authorized institution such as NPCI or a reputed private institution.

With δ-Money clients can get suspended if they engage in any malicious transactions. In the event, any malicious transaction is detected its impacts can be reversed and safeguards put in the mechanism of exiting the system by exchanging δ-Money with some other form of money. The possibilities of collusion between parties are rather limited as unless all peers report consistent facts, they run afoul of integrity checks.

Three are many extension possibilities. The transactions can be colour-coded to predict any anomalies. For bank loans, we can build the capability to track follow-on transactions. DDCS support to Data Clients is particularly useful here.

Finally, δ-Money fulfills all three functions of a currency adequately as the medium of exchange, store of value, and unit of account. As it is issued by a bank, there should be no issue in using it as a medium of exchange. It does a particularly good job as a store of value, as the monies with the bank cannot be lent and the valuation can remain stable compared to both fiat currencies and crypto-currencies. δ-Money serves well as a unit of account as it can do a good job of measuring the economic value of any item as any fiat currency. δ-Money can handle regulatory issues such as account insurance, taxation, and illegal usage with appropriate configuration and customization. Unlike bitcoin and physical currencies, the owners can be tracked easily. At the same time, any information can be shared only on a need-to-know basis. By restricting δ-Money transactions only to national/regional jurisdiction Governments can avoid the risk of flight of money.

δ-Money is designed to avoid hoarding which both physical currency and crypto-currencies such as bitcoin are prone to. A payment bank can limit the amounts in accounts as well as amounts per transaction. Thus, a Government can balance the two sets of economies intelligently. One of consumption and another of savings and investment.

δ-Money however is not suitable for environments where transactions are completely random. This means a peer is not likely to repeatedly transact with the same peer. Then it will continue to hold the state for too many peers without particular value.

DDCS also assumes that Currency Clients generally take care of their devices and client applications are protected from cybersecurity attacks. We can introduce additional steps where Currency Manager runs health checks on Currency Clients periodically. Adding new peers can have a higher security bar. We can introduce processes such as parties digitally signing their transaction statements every month. Further, the failure of the Currency Manager and Integrity Manager can put the DDCS in limbo, requiring some housekeeping function to be discharged continually as in any IT system.

From a distributed computing perspective, DDCS ensures eventual consistency, liveness, correctness, and fairness. From a storage/information management perspective, each Currency Client stores its transactions in a B+ Tree which provides a very efficient mechanism to store data and support queries.

# 5. India Considerations

We believe the Indian citizens would value a new alternative in the form of δ-Money which gives them the ability to transact in an environment of mutual trust with a strong focus on data integrity. At a macro level, δ-Money can lead to greater financial stability, increase contestability in retail transactions, address financial inclusion, and prevent frauds.

Even though the volume of digital payments has exploded in India, many institutions find it hard to reconcile payments with the parties who paid. The peer-to-peer model which mandates peer registration with mutual consent can be particularly useful here, where an institution or merchant can associate a customer number/consumer number/student or employee enrolment number with such registration and they can get reports on payments made as needed.

The approach used by DDCS is in philosophical consonance with that of edge computing where clients hold data, processing is done as close to data as possible and only what is needed in transmitted over the network, while retaining significant connectivity among clients themselves. This along with emergence of 5G networks that can connect billions of devices can make DDCS an attractive proposition for India. We also expect newer devices that are lot more secure and powerful to emerge which can be particularly useful for financial transactions, that even common people can afford.

Bank of Canada Study Paper [33] breaks down the complexity of currency management elegantly by classifying monies there-fold: Outside Money (Physical Currency, Digital Currency) which is a claim on Central Bank, Inside Money (Deposit Accounts) is a claim backed by private credit (banks can create money through fractional reserve lending) and helicopter money (direct transfer of funds to individuals and firms by quantitative easing i.e., by running fiscal deficits). Under this classification δ-Money is essentially an Outside Money.

The DDCS can also be used to manage Helicopter Money. This can be done by managing the Social Sector spending on the DDCS platform. This can give a greater ability for Government to track the benefits directly transferred to target groups. This can be done by concerned Government Departments operating as Data Clients in DDCS. Further DDCS can also be used to manage loan accounts by giving banks the ability to be informed of follow-on transactions that stem from loans disbursed by them. Data Client role comes in handy here as well. Similar capability can be used by large corporate and government departments to manage their routine expenses which involve known vendors and service providers. DDCS system can be enhanced to gather insights to prevent frauds as well as opportunities to optimize the spending. Preventing frauds by itself can bring immense value to the Indian Economy considering routine reportage on misuse of loans and subsidies [60].

# 6. Conclusion

In this paper, we have proposed a Decentralized Digital Currency System (DDCS) that manages. DDCS is architected with separation of concerns as a guiding principle. where Currency Manager's concern is Currency Conservation, Integrity Manager's concern is consistency of transaction accounts and Currency Client's concern is to transact with other peers and ability to keep the state of transactions it engages with others in self-custody. Clients register themselves as peers and work based on Mutual Consensus and commit to transactions. Use of Authenticated Data Structure enables them to transact where Currency Manager and Integrity Manager act as verifiers and raise alerts or suspend clients only in case of anomalies. Otherwise, the Clients can commit the transactions in a disintermediated manner. Verification happens post-facto and does not come in the way of transaction. The sweet-spot for δ-Money is routine, repeated and periodic transaction with same counter-parties which is likely in our personal and professional lives. We recommend that Payment Banks issue δ-Money and manage its valuation. There are added benefits of using DDCS. By acting as Data Clients, Banks, Corporates and Governments can track the utilization of operating expenses, loans or benefits and validate their legitimate use subject to creating an appropriate legal framework to do that. Research is required to assess possible cyber-security threats to DDCS and design suitable mechanisms to address them.

# Declarations

## Availability of data and material

Not applicable.

## Competing/Conflicting interests

None.

Please note that the method proposed in this paper is disclosed in the Indian Patent Application no. 201941044512 published on December 21, 2019 at a broad level. The said method has been further refined and elaborated in this paper.

## Funding

None

## Authors' contributions

First Author responsible for overall conception and construction of the article

Second Author responsible for strengthening the literature review and overall technical quality of article.

Third and Fourth authors contributed to building an early version of prototype as part of their Undergraduate Projects under the guidance of first author.

## Acknowledgements

Authors acknowledge the affiliating institutions for the support extended to pursue this research.

52. Yongzhi Wang, Yulong Shen, Hua Wang, Jinli Cao, and Xiaohong Jiang, MtMR: Ensuring MapReduce Computation Integrity with Merkle Tree-Based Verifications, IEEE Transactions On Big Data, Vol. 4, No. 3, July-September 2018
53. J. -F. Pâris and T. Schwarz, "Merkle Hash Grids Instead of Merkle Trees," 2020 28th International Symposium on Modeling, Analysis, and Simulation of Computer and Telecommunication Systems (MASCOTS), 2020, pp. 1-8, DOI: 10.1109/MASCOTS50786.2020.9285942.
54. Y. -q. Fang, J. -b. Liao and L. -y. Lai, "Verifiable Secret Sharing Scheme Using Merkle Tree," *2020* International Symposium on Computer Engineering and Intelligent Communications (ISCEIC)*,* 2020, pp. 1-4, DOI: 10.1109/ISCEIC51027.2020.000*08.*
55. Adi Shamir, How to share a secret, Communications of the ACM, November 1979, https://doi.org/10.1145/359168.359176
56. Jian Xu, Laiwen Wei, Wei Wu, Andi Wang, Yu Zhang, Fucai Zhou, Privacy-preserving data integrity verification by using lightweight streaming authenticated data structures for healthcare cyber-physical system, Future Generation Computer Systems, Volume 108,2020, Pages 1287-1296, ISSN 0167-739X, https://doi.org/10.1016/j.future.2018.04.018
57. Mohammad Etemad, Alptekin Kupcu, Database Outsourcing with Hierarchical Authenticated Data Structures**.** IACR Cryptol. ePrint Arch. 2015: 351 (2015)
58. D. Pennino, M. Pizzonia and A. Papi, "Overlay Indexes: Efficiently Supporting Aggregate Range Queries and Authenticated Data Structures in Off-the-Shelf Databases," in IEEE Access, vol. 7, pp. 175642-175670, 2019, doi: 10.1109/ACCESS.2019.2957346
59. Jensen C.S., Snodgrass R.T. (2009) Temporal Database. In: LIU L., ÖZSU M.T. (eds) Encyclopedia of Database Systems. Springer, Boston, MA. https://doi.org/10.1007/978-0-387-39940-9_395
60. Yatish Yadav, Forty of 50 cooperative banks in Uttar Pradesh hit by fraud; loan amounts forged, subsidy money siphoned off, Firstpost, June 26, 2019.:
Page **37** of **37**